\documentclass
[floatfix,superscriptaddress,secnumarabic,amssymb,amsmath,nobibnotes,aps,prd,showkeys,nofootinbib,onecolumn,notitlepage,10pt]{revtex4}%
\usepackage{setspace}
\usepackage{color}
\usepackage{amsmath}
\usepackage{amsfonts}
\usepackage{verbatim}
\usepackage{amssymb}
\usepackage{graphicx,bm}
\usepackage{graphicx}
\usepackage{amsmath}
\usepackage{amssymb}
\usepackage{amssymb}
\usepackage{graphicx,bm}
\usepackage{graphicx}
\usepackage[colorlinks]{hyperref}
\usepackage[caption=false]{subfig}%
\hypersetup{
breaklinks=true,
pdfstartview={FitH},    
colorlinks=true,       
linkcolor=blue,          
citecolor=red,        
filecolor=magenta,      
urlcolor=blue,           
anchorcolor=green,      
linktocpage=true
}
\providecommand{\U}[1]{\protect\rule{.1in}{.1in}}

\newcommand{\be}{\begin{equation}}
\newcommand{\ee}{\end{equation}}

\newcommand{\mincir}{\raise
-3.truept\hbox{\rlap{\hbox{$\sim$}}\raise4.truept\hbox{$<$}\ }}
\newcommand{\magcir}{\raise
-3.truept\hbox{\rlap{\hbox{$\sim$}}\raise4.truept\hbox{$>$}\ }}

\ifx\pdfoutput\relax\let\pdfoutput=\undefined\fi
\newcount\msipdfoutput
\ifx\pdfoutput\undefined\else
\ifcase\pdfoutput\else
\ifx\paperwidth\undefined\else
\ifdim\paperheight=0pt\relax\else\pdfpageheight\paperheight\fi
\ifdim\paperwidth=0pt\relax\else\pdfpagewidth\paperwidth\fi
\fi\fi\fi

\begin{document}
\title{Viaggiu holographic dark energy in light of DESI DR2}
\author{Amlan K. Halder}
\email{amlankanti.halder@woxsen.edu.in}
\affiliation{School of Sciences, Woxsen University, Hyderabad 502345, Telangana, India}
\author{Andronikos Paliathanasis}
\email{anpaliat@phys.uoa.gr}
\affiliation{Departamento de Matem\`{a}ticas, Universidad Cat\`{o}lica del Norte, Avda.
Angamos 0610, Casilla 1280 Antofagasta, Chile}
\affiliation{Institute of Systems Science, Durban University of Technology, Durban 4000,
South Africa}
\affiliation{Centre for Space Research, North-West University, Potchefstroom 2520, South Africa}
\affiliation{National Institute for Theoretical and Computational Sciences (NITheCS), South Africa}
\affiliation{School of Sciences, Woxsen University, Hyderabad 502345, Telangana, India}
\author{Stefano Viaggiu}
\email{s.viaggiu@unimarconi.it}
\affiliation{Dipartimento di Scienze Ingegneristiche, Universit\'a degli Studi Guglielmo Marconi, Via Plinio 44, I-00193 Roma, Italy}
\author{Abdulla Al Mamon}
\email{abdulla.physics@gmail.com}
\affiliation{Department of Physics, Vivekananda Satavarshiki Mahavidyalaya (affiliated to the Vidyasagar University), Manikpara-721513, West Bengal, India}
\author{Subhajit Saha}
\email{subhajit1729@gmail.com}
\affiliation{Department of Mathematics, Panihati Mahavidyalaya, Kolkata 700110, West Bengal, India}

\begin{abstract}
We test the cosmological viability of the Viaggiu holographic dark energy (VHDE) model by using late-time observational data. In particular, we place constraints on the free parameters of the model using Type Ia supernovae from the PantheonPlus, Union3.0, and DES-Dovekie catalogues, the Cosmic Chronometers, and the Baryon Acoustic Oscillations from the DESI DR2. Our analysis suggests that the VHDE model fits the observational data better or similar to the $\Lambda$CDM for all dataset combinations considered. The value obtained for $H_0$ is similar to the $\Lambda$CDM, while the current matter density parameter is constrained around $\Omega_{m0}\simeq 0.24$, smaller to that obtained by the $\Lambda$CDM. Moreover, the parameter introduced by the VHDE is found to have a mean value within the range $\frac{\pi}{3} \delta^2 \sim 0.27-0.33$. Finally, we used Akaike's Information Criterion (AIC) and Bayesian evidence to test the VHDE model against the $\Lambda$CDM scenario. The AIC demonstrates that the two models are statistically indistinguishable, while Bayesian evidence reveals that the data have a mild preference for the $\Lambda$CDM model for most of the dataset combinations considered. Nevertheless, the VHDE model remains consistent with current late-time cosmological observations and offers a feasible mechanism for describing the late-time accelerating scenario. 
\end{abstract}

\date{\today}
\maketitle

\section{Introduction}
\label{sec1}
It is evident from observational probes~\cite{acc1,acc2,acc3,acc4,acc5} that our cosmos is experiencing an accelerated expansion at present. Several models have been proposed in the literature to explain this unexpected cosmic phenomenon. Nevertheless, none of them have been found to be fully consistent with all astronomical observations available to us. Cosmologists have attributed this accelerated expansion to a mysterious force referred to as ``dark energy.'' This phrase was first introduced in a paper by Bahcall et al. \cite{Bahcall1}. Among all these models, the $\Lambda$-cold-dark-matter ($\Lambda$CDM) model has consistently been shown to have an edge over other models, as it exhibits excellent alignment across a broad spectrum of observational datasets. Consequently, the $\Lambda$CDM model has come to be broadly acknowledged as the standard model of Cosmology. The $\Lambda$CDM framework is characterized by a small cosmological constant $\Lambda$, which represents the dark energy (DE) candidate, and cold (non-relativistic) dark matter (DM) in the form of pressureless dust, which acts as the gravitational ``glue'' that allows the formation of cosmic structures and binds the galaxies together. Observational data reveal that these two components constitute nearly 96\% of the total energy in the Universe. Although the analysis of data obtained with modern telescopes continues to support the $\Lambda$CDM model, the origin and the underlying nature of DE are not yet fully understood. Moreover, $\Lambda$ suffers from several fundamental challenges, such as the cosmic coincidence problem and the fine-tuning problem \cite{refft1,refccp1,Peebles1,Padmanabhan1}.\\

Over the past few years, cosmologists have also identified some key challenges for the $\Lambda$CDM model, particularly the $H_0$ and $S_8$ tensions \cite{refh03,refh01,refs8,refh02}. Furthermore, numerical simulations based on the $\Lambda$CDM model predict several phenomena occurring at galactic and sub-galactic scales that appear to be inconsistent with observations. These issues include the missing satellite, core-cusp, and the ``too big to fail'' problems, and are collectively called ``small-scale-Cosmology crisis.'' The interested readers may consult the nice reviews, cf. \cite{Perivolaropoulos1,Mavromatos1} to get an overview of these persistent discrepancies within the standard $\Lambda$CDM model. It is interesting to note that an array of alternative scenarios have been proposed and studied in detail in the literature, but none of these models provides a clear advantage over the $\Lambda$CDM model upon comparison with observational data \cite{Copeland1,Frieman1,Caldwell1,Silvestri1,Li0}.\\

Applying the holographic principle (HP) to the dark energy problem has led to the development of the holographic dark energy (HDE) paradigm as an alternative approach. First conceived by G. 't Hooft \cite{tHooft1}, the HP asserts that the degrees of freedom enclosed within a volume of space can possibly be visualized as a hologram, which gives a theory formulated on the boundary of the volume. It was Susskind \cite{Susskind1} who later presented an interpretation of HP in the context of string theory. Interestingly, the AdS/CFT correspondence, conjectured by Maldacena \cite{Maldacena1} in 1999, is widely accepted to be the most successful realization of HP. Li was the first \cite{Li1} to introduce the first HDE model in Cosmology. Using HP, Li expressed the energy density of DE as proportional to the product of the square of the reduced Planck mass and the inverse square of the characteristic length scale. When the Hubble and particle horizons were chosen as the length scales, the late-time cosmic acceleration could not be reproduced. Subsequently, Li chose the future event horizon as the length scale, and the model showed satisfactory agreement with the contemporary observational data. This led the HDE model to be accepted as a promising alternative to the $\Lambda$CDM model. Moreover, HDE scenarios have been shown to be free from potential difficulties that are likely to arise in modified gravity theories \cite{Akrami1}.\\

A wide range of HDE models have been proposed simply by altering the functional form of the underlying entropy. Some of the well-studied modified entropies include the nonadditive entropy proposed by Tsallis \cite{Tsallis1}, the relativistic entropy proposed by Kaniadakis \cite{Kaniadakis1,Kaniadakis2}, power-law corrected entropies \cite{Das1,Radicella1}, and the quantum-gravity-corrected entropy proposed by Barrow \cite{Barrow1}. We obtain, respectively, Tsallis \cite{Saridakis0,Sadri1}, Kaniadakis \cite{Drepanou1,Almada1,Luciano1}, power-law \cite{Telali1}, and Barrow \cite{Saridakis1} HDE models, from these modified entropies. These modified entropy formalisms contain a parameter that characterizes the departure from the standard Bekenstein-Hawking entropy. The above models, together with some lesser-known HDE scenarios, notably the R\'enyi \cite{Moradpour1} and Sharma-Mittal \cite{Jahromi1} HDE models, have also been studied with various length scales; see, for example, the recent work \cite{Luciano2}, and in the presence of DM interactions. For a comprehensive overview of the HDE framework and its modifications, one may refer to the review by Wang, Wang, and Li \cite{Wang1}. As a closing remark, it is worth mentioning that recently a generalized form of horizon entropy has been proposed that establishes a correspondence between the equations of a FLRW universe in a general theory of gravity and the thermodynamics of the dynamical apparent horizon \cite{Nojiri1,Nojiri2,Odintsov1,Luciano3,Luciano:2026ufu,Leizerovich:2026pfy}. Most of the entropy forms discussed above can be deduced as special cases from this generalized entropy.\\ 

This paper examines the Viaggiu holographic dark energy (VHDE) as a mechanism for describing the late-time accelerated expansion of the Universe. This HDE model has recently been introduced by Saha et al. \cite{5} using a generalized form of the Bekenstein-Hawking entropy presented in 2014 by Viaggiu \cite{3,4}. The authors consider Hubble and future event horizons as infrared (IR) cut-offs in their study. When no interaction between DE and DM is assumed, the Hubble horizon yields a dusty universe, however, when interaction between the two components is introduced, it demonstrates an Einstein static universe. However, when the future event horizon is considered as the IR cut-off, the noninteracting VHDE model gives a very good description of the late-time evolutionary behavior of the Universe, and the deductions are consistent with recent results obtained with the standard HDE model investigated using a similar characteristic length scale \cite{Tang1,Mandal1}. \\  

For our analysis, we combine observational data from multiple sources. In particular, we consider measurements of the Baryon Acoustic Oscillations (BAO) from the DESI DR2, together with Cosmic Chronometers, and one of three Type Ia Supernova catalogues: PantheonPlus, Union3.0, or the recent DES-Dovekie dataset. These datasets probe the expansion history of the late Universe at different distances. The consideration of multiple SNIa catalogs helps to check consistency and reduces biases from different data processing.\\ 

The remainder of the paper is structured as follows. We introduce the basic elements of the VHDE model in Section \ref{sec2}. The modified cosmological equations of the model are presented in Section \ref{sec3}. Section \ref{sec4} discusses the results of our study, where we perform the observational tests for the VHDE model. We present our conclusions in Section \ref{sec5}.

\section{Viaggiu entropy and the VHDE model}
\label{sec2}
The Bekenstein argument \cite{1} states that in an asymptotically flat spacetime there exists an universal upper bound ($S_{max}$) on the entropy $S$ of a spherical region of areal radius $R$ and energy $E$ given by
\begin{equation}
S\leq S_{max}=\frac{2\pi k_B R E}{c\hbar}.
\label{a}
\end{equation}
Since \cite{Susskind1} the maximum energy $E_{max}$ allowed in the aforementioned  spherical region is the one provided by the largest black hole (BH) that can fit inside $R$, we have $E_{max}=Mc^2=c^4R/(2G)$, where $M$ is the ADM mass of the BH. Therefore, from (\ref{a}) we obtain the well known Bekenstein-Hawking entropy formula
$S_{BH}=S_{max}=k_B\frac{A_h}{4L_p^2}, A_h=4\pi R^2$.\\

In an expanding universe, the situation is quite different. In fact, the entropy of a given region is indicative of the information, or the content of energy, inside. Moreover, the expansion of the Universe makes the formation of BHs more difficult, requiring more localized energy than in the static case. Obviously, more energy does imply more entropy. To quantify the aforementioned surplus of entropy corresponding to the static case, in \cite{3,4}, theorems concerning the formation of trapped surfaces are used. In particular, in a Friedmann flat context, if for a sphere $\Sigma$ with proper radius $L$, proper area $A=4\pi L^2$ and with a pure-trace extrinsic curvature a mass excess $\delta M$ given by
\begin{equation}
\delta M \frac{G}{c^2}<\frac{L}{2}+\frac{AH}{4\pi c}
\label{b}
\end{equation}
is present, then $\Sigma$ is not trapped. After using the Bekenstein bound (\ref{a}) with $E_{max}=c^2\delta M$
and $R=L$
we arrive to the expression for the entropy $S_{\Sigma}$ within $\Sigma$:
\begin{equation}
S_{\Sigma}=k_B\frac{A}{4L_p^2}+\frac{3k_B}{2cL_p^2}VH,
\label{c}
\end{equation}
where $H$ is the Hubble flow. In \cite{3,4}, the entropy (\ref{c}) has been used to study the expression of the first law of thermodynamics at the apparent horizon 
$L_h$. In \cite{5}, expression (\ref{c}) has been used to build a new HDE model (VHDE).
To this purpose, the starting point is the expression for the upper bound $\rho_d$ of the quantum zero-point energy density arising from an ultraviolet (UV) cut-off. More precisely, by following \cite{1,Li1,Wang1}, $\rho_d$ in a region of proper side $L$ should not exceed the maximum mass of a BH sitting inside $L$. In formulas, we have:
\begin{equation}
\frac{L^4\rho_d}{M_p^2}\leq S,
\label{d}
\end{equation}
where $M_p=\sqrt{\frac{c\hbar}{8\pi G}}$ and $S$ denotes the entropy. After using formula (\ref{c})
with $S_{\Sigma}=S$ in expression (\ref{d}) and by using geometric units with $G,c,\hbar, k_B$ equal
to 1, we obtain \cite{5}
\begin{equation}
S=\pi L^2+2\pi L^3 H.
\label{e}
\end{equation}
After considering inequality (\ref{d}) saturated with entropy formula (\ref{e}) and the largest $L$
allowed, we get
\begin{equation}
\rho_d=\frac{{\delta}^2}{8\pi}L^{-4}\left(\pi L^2+2\pi H L^3\right),
\label{f}
\end{equation}
where ${\delta}$ is a positive numerical constant (see \cite{5} for more details). \\

The expression (\ref{f}) for HDE has been firstly applied in \cite{5} with $L$ the Hubble horizon $L=1/H$. To this purpose, the results in \cite{5} show that the VHDE model with the IR cut-off given by the apparent horizon in any case is not capable of explaining current cosmological data. Attention is thus shifted on models with the future event horizon as IR cut-off. In fact, from the  literature it seems that these HDE models can reproduce late-time cosmic acceleration with a good fit with observational data (see for example \cite{5,8} and references therein).\\ 

At this point, a comparison with other HDE models considered recently is necessary. First of all, we stress that Viaggiu entropy formula (5) contains the Bekenstein-Hawking one only in the limit $H=0$, i.e. in the static case where dynamical degrees of freedom related to the expansion of the Universe are absent. In fact, Viaggiu entropy is the sum of the Bekenstein Hawking formula (nonextensive) with a volume term (extensive term) and consequently Viaggiu entropy is nonextensive. Moreover, Viaggiu entropy is physically well-motivated by theorems of general relativity regarding BH formations in expanding Friedmann universes. In Ref. \cite{a}, the authors consider the so called Barrow $\Delta$ entropy and Tsallis-Cirto $\delta$ entropy (see Ref. \cite{a} and references therein). The HDE density $\rho_{HDE}$ in \cite{a} looks like $\rho_{HDE}\sim S(L)L^{-4}$: the $\rho_{d}$ considered in this paper can be obtained by replacing $S(L)$ with expression (5), thus obtaining the VHDE density given in Eq. (6). Moreover, in Ref. \cite{a}, $S(L)$ is chosen in the case of Barrow entropy to be $S_B(L)\sim L^{1+\Delta}$ where $\Delta$ is a real parameter with $0\leq \Delta \leq 1$. The Tsallis-Cirto entropy is obtained with  $\Delta=2(\delta-1)$. Note that although with $\Delta=1$ or $\delta =3/2$ we obtain an extensive Barrow/Tsallis-Cirto entropy scaling as the volume, we cannot get expression (5) for any value of $\Delta$. Stated in other words, the Viaggiu entropy (5) can be obtained only by summing two Barrow entropies, i.e $S_{B_1}(\Delta=0)+S_{B_2}(\Delta=1)$. To this regard, Viaggiu entropy is new with respect to the ones considered in Ref. \cite{a}. A similar reasoning applies to the proposals in Ref. \cite{b} and Ref. \cite{c}. In Ref. \cite{b}, the following entropy $S(L)$ is considered with $S(L)\sim {(1+k_1 S(L))}^{\beta}-{(1+k_2 S(L))}^{-\beta}$, with $\{k_1, k_2,\beta\}$ positive constants and $S(L)$ the Bekenstein-Hawking one. Also in this case, it is easy to see that combination (5) cannot be obtained. Finally, in Ref. \cite{c},  in order to obtain a finite entropy at the bounce $H(t)=0$, $S(L)$ given by the Bekenstein-Hawking entropy is substituted with $\tanh (S(L))$ and also in this case the expression (6) cannot be obtained. Summarizing, the VDHE model is a physically motivated cosmological model that is structurally distinct from other entropic models, and not yet studied in current literature as in Refs. \cite{a,b,c}.\\

In the next section, we present the main formulas for the VHDE model with IR future horizon cut-off that  will be used to test the model against current data.

\section{Main formulas in the VHDE model with future event horizon as the IR cut-off}
\label{sec3}
The future event horizon, which exists only in an accelerating universe, is defined mathematically as follows \cite{Faraoni1}
\begin{equation} \label{re-def}
R_E=a(t)\int_{t}^{\infty} \frac{dt}{a(t)}=a\int_{a}^{\infty} \frac{da}{Ha^2}=a\int_{x}^{\infty} \frac{dx}{Ha}.
\end{equation}
From a physical perspective, the future event horizon represents the proper distance to the farthest event that can ever be observed by a comoving observer \cite{Faraoni1}.\\

Now, following \cite{5}, if we put $L=R_E$ in Eq. (\ref{f}), the energy density of VHDE becomes
\begin{equation}\label{rhod-re}
\rho_d=\frac{\delta^2}{8R_{E}^2}(1+2HR_E).
\end{equation}
Again, when DE and DM do not interact, the equation of state (EoS) of DE is given by
\begin{equation} \label{wd-new}
w_d=-\frac{\Omega_{d}^{'}}{3\Omega_d(1-\Omega_d)},
\end{equation}
where $\Omega_{d}$ is the fraction of the total energy density associated with DE.\\ 

It is important to determine $\Omega_{d}^{'}$ in order to understand $w_d$. To this effect, we first compute the derivative of the VHDE density with respect to cosmic time \cite{5}:
\begin{equation}
\dot{\rho}_d=\frac{\delta^2}{4R_{E}^{3}}\left[1+R_{E}^{2}(\dot{H}-H^2)\right].
\end{equation}
In addition, we have the time-derivative of $\Omega_d$ as \cite{5}
\begin{equation}
\dot{\Omega}_{d} =\frac{2\pi \delta^2}{3H^2R_{E}^{2}}\left[-R_E(\dot{H}+H^2)+\frac{1}{R_E}-\frac{\dot{H}}{H}\right].
\end{equation}
It is more convenient to consider the above derivative with respect to $\mbox{ln}~a$, which gives 
\begin{equation}\label{omd-dash}
\Omega_{d}^{'}=\frac{2\pi \delta^2}{3H^2R_{E}^{2}}\left[-R_E\frac{\dot{H}}{H}-HR_E+\frac{1}{HR_E}-\frac{\dot{H}}{H^2}\right].
\end{equation}
Introducing $\Omega_{d}^{'}$ in Eq. (\ref{rhod-re}) leads to the quadratic equation \cite{5}
\begin{equation} \label{quad-e}
(HR_{E})^2-\left(\frac{2 \pi \delta^2}{3\Omega_{d}}\right)HR_E-\frac{\pi \delta^2}{3\Omega_{d}}=0.
\end{equation}
Following the argument in \cite{5}, we obtain
\begin{equation}\label{hre}
HR_E=\frac{\pi \delta^2}{3\Omega_d}\left(1+\sqrt{1+\frac{3\Omega_d}{\pi \delta^2}}\right).
\end{equation}
Then, from Eq. (\ref{re-def}), and the Friedmann equation, in which $\rho_m$ is the energy density of DM, which is assumed to be in the form of pressureless dust, that is,
\begin{equation}
3H^2=8\pi (\rho_m+\rho_d),   
\end{equation}
we finally obtain \cite{5}
\begin{eqnarray} \label{omdd}
\frac{\Omega_{d}^{'}}{\Omega_{d}(1-\Omega_{d})}=\frac{\frac{6\Omega_d}{\pi \delta^2}+\left(1+\sqrt{1+\frac{3\Omega_d}{\pi \delta^2}}\right)}{(2-\Omega_d)\left(1+\sqrt{1+\frac{3\Omega_d}{\pi \delta^2}}\right)-\frac{\frac{3\Omega_d}{\pi \delta^2}(1-\Omega_d)}{\sqrt{1+\frac{3\Omega_d}{\pi \delta^2}}}}.
\end{eqnarray}
Furthermore, on multiplying Eq. (\ref{omdd}) by $-\frac{1}{3}$, the EoS of VHDE is obtained as \cite{5}
\begin{eqnarray} \label{wd-f}
w_d &=& -\frac{\Omega_{d}^{'}}{3\Omega_{d}(1-\Omega_{d})} \nonumber \\
&=& -\frac{1}{3}\left[\frac{\frac{6\Omega_d}{\pi \delta^2}+\left(1+\sqrt{1+\frac{3\Omega_d}{\pi \delta^2}}\right)}{(2-\Omega_d)\left(1+\sqrt{1+\frac{3\Omega_d}{\pi \delta^2}}\right)-\frac{\frac{3\Omega_d}{\pi \delta^2}(1-\Omega_d)}{\sqrt{1+\frac{3\Omega_d}{\pi \delta^2}}}}\right].
\end{eqnarray}
Moreover, the deceleration parameter $q$ in the VHDE model is computed as \cite{5}
\onecolumngrid
\begin{eqnarray}
q &=& \frac{1}{2}(1+3w_d \Omega_d) \nonumber \\
&=& \frac{1}{2}-\frac{\Omega_d \sqrt{1+\frac{3\Omega_d}{\pi \delta^2}} \left[\left(1+\frac{3\Omega_d}{\pi \delta^2}\right)^{\frac{3}{2}}+\left(1+\frac{9\Omega_d}{2\pi \delta^2}\right)\right]}{\left(1+\sqrt{1+\frac{3\Omega_d}{\pi \delta^2}}\right)\left[\frac{3\Omega_d}{\pi \delta^2}+(2-\Omega_d)\left(1+\sqrt{1+\frac{3\Omega_d}{\pi \delta^2}}\right)\right]}.
\end{eqnarray}

\section{Observational Data Analysis}
\label{sec4} 
The observational constraints are reported in this Section. In
the following, we summarize the data sets considered in this work and the
methodology adopted for the statistical analysis.

\subsection{Observational Data}

For our analysis, we employ cosmological data of the late-universe which allow
us to constraint the evolution of the background geometry. In particular we
consider Supernova data (SNIa) as provided by three different catalogues, the
supernova Type Ia (SNIa) data, the Baryonic Acoustic Oscillations (BAO) from the DESI DR2 collaboration, and the Cosmic Chronometers (CC).

\begin{itemize}
\item SNIa: We utilize three supernova Type Ia catalogues in this analysis:
PantheonPlus (PP) \cite{Brout:2022vxf}, Union3.0 (U3) \cite{rubin2023union}
and DES-Dovekie (DESD) \cite{DES:2025sig}. Each catalogue provides
measurements of the distance modulus $\mu^{obs}$ as a function of the redshift.
The PP dataset comprises 1550 SNIa observations at $10^{-3}<z<2.27$. The U3
compilation contains 2087 events covering a comparable redshift range. While
these two catalogues share 1363 common SNIa events, they employ different
photometric reduction pipelines, resulting in independent datasets despite overlap. The DESD catalogue presents a reanalysis of the five-year Dark Energy
Survey supernova program (DES-SN5YR), which yielded 1820 SNIa measurements at lower
redshifts ($z<1.13$)

\item BAO: Our analysis incorporates baryon acoustic oscillation measurements
from the Dark Energy Spectroscopic Instrument Data Release 2 (DESI DR2)
\cite{DESI:2025zpo,DESI:2025zgx,DESI:2025fii}. The DESI DR2 dataset provides
observables at seven redshift bins, each normalized by the sound horizon at
the baryon drag epoch $r_{drag}$. The observables are the transverse comoving
angular distance ratio, $\frac{D_{M}}{r_{drag}}=\frac{D_{L}}{\left(
1+z\right)  r_{drag}},~$the volume-averaged distance ratio, $\frac{D_{V}%
}{r_{drag}}=\frac{\left(  zD_{H}D_{M}^{2}\right)  ^{1/3}}{r_{drag}}$ and the
and the Hubble distance ratio $\frac{D_{H}}{r_{d}}=\frac{c}{r_{drag}H(z)}.$

\item CC: We employ 31 model-independent measurements of the Hubble parameter
$H\left(  z\right)  $ derived from the cosmic chronometer method
\cite{cc1}~spanning $0.09\leq z\leq1.965.~$By measuring the age difference
$\frac{dz}{dt}$ between galaxies at neighboring redshifts, the Hubble
parameter is obtained directly as $H\left(  z\right)  =-\frac{1}{\left(
1+z\right)  }\frac{dz}{dt}$. The likelihood analysis incorporates the entire
covariance matrix\footnote{https://gitlab.com/mmoresco/CCcovariance} following
the methodology described in \cite{Moresco:2020fbm}.
\end{itemize}

For our analysis, we consider six different combinations of the datasets, that
is, \ (I) PP\&BAO, (II)\ PP\&CC\&BAO, (III) U3\&BAO, (IV)\ U3\&CC\&BAO, (V)
DD\&BAO, and (VI)\ DD\&CC\&BAO.

\subsection{Methodology \& Priors}

We perform Bayesian parameter estimation using the COBAYA
framework\footnote{https://cobaya.readthedocs.io/} \cite{cob1,cob2} with a
custom theory module to compute the cosmological background evolution
numerically with the Runge-Kutta algorithm. We make use of the PolyChord
nested sampling algorithm \cite{poly1,poly2,getd} which provides the posterior
distributions and the Bayesian evidence. \\

For the VHDE model, the free parameters are $\left\{  H_{0},\Omega
_{m0},r_{drag},\zeta\right\}  $, with $\zeta=\frac{\pi}{3}\delta^{2}$.
In addition, we perform the same observational tests and for the $\Lambda$CDM
which is used as the reference model. The dimension of the parameter space for
the $\Lambda$CDM is three, consisting of the variables $\left\{  H_{0}%
,\Omega_{m0},r_{drag}\right\}  $. The priors applied in this analysis are
presented in Table \ref{prior}. \\

\begin{table}[tbp] \centering
\caption{Priors of the Free Parameters for the PolyChord sampler.}%
\begin{tabular}
[c]{ccc}\hline\hline
\textbf{Priors} & \textbf{VHDE} & $\Lambda$\textbf{CDM}\\\hline
$\mathbf{H}_{0}$ & $\left[  60,80\right]  $ & $\left[  60,80\right]  $\\
$\mathbf{\Omega}_{m0}$ & $\left[  0.1,0.5\right]  $ & $\left[  0.1,0.5\right]
$\\
$\mathbf{r}_{drag}$ & $\left[  120,170\right]  $ & $\left[  120,170\right]
$\\
$\mathbf{\zeta}~\left(  \frac{\pi}{3}\delta^{2}\right)  $ & $(0,2]$ &
$-$\\\hline\hline
\end{tabular}
\label{prior}
\end{table}

For the statistical comparison of the above two models, we employ the Akaike
Information Criterion (AIC) \cite{AIC} and Bayesian evidence
\cite{AIC2}\thinspace. The parameter $AIC$ is calculated using the $\chi_{\min}^{2}$ and
the dimension $\mathcal{N}$ of the parametric space with the algebraic
formula
\begin{equation}
AIC=\chi_{\min}^{2}+2\mathcal{N},
\end{equation}

The AIC provides a statistical criterion to determine the preferred model among different candidates, that is, of $\left\vert \Delta AIC\right\vert =\left\vert
AIC_{QC}-AIC_{\Lambda}\right\vert $. Specifically, if $\left\vert \Delta
AIC\right\vert <2$, both models are statistically indistinguishable, while for
$\left\vert \Delta AIC\right\vert <6$, the model with the smaller $AIC$ is weakly preferred. On the other hand, if $\left\vert \Delta
AIC\right\vert >6$ the model under consideration is strongly disfavored. \\

Nevertheless, Jeffrey's scale \cite{AIC2} provides a scale for the comparison
of two models using the Bayesian evidence values.
Specifically, we calculate the difference of the Bayesian evidence, $\Delta\left(  \ln Z\right)  =\ln\frac{Z_{1}}{Z_{2}}$. According
to Jeffrey's scale if $\left\vert \Delta\left(  \ln Z\right)  \right\vert <1$,
the two models are statistically comparable. When $\left\vert \Delta\left(  \ln Z\right)
\right\vert >1$, the model with the larger Bayesian evidence is weakly favored. If \ $\left\vert \Delta\left(  \ln Z\right)
\right\vert >2.5$, the evidence becomes moderate, while for $\left\vert
\Delta\left(  \ln Z\right)  \right\vert >5$, there is decisive evidence that the dataset supports the model with the higher Bayesian evidence.

\subsection{Results}%
\begin{figure}[h]
\centering\includegraphics[width=1\textwidth]{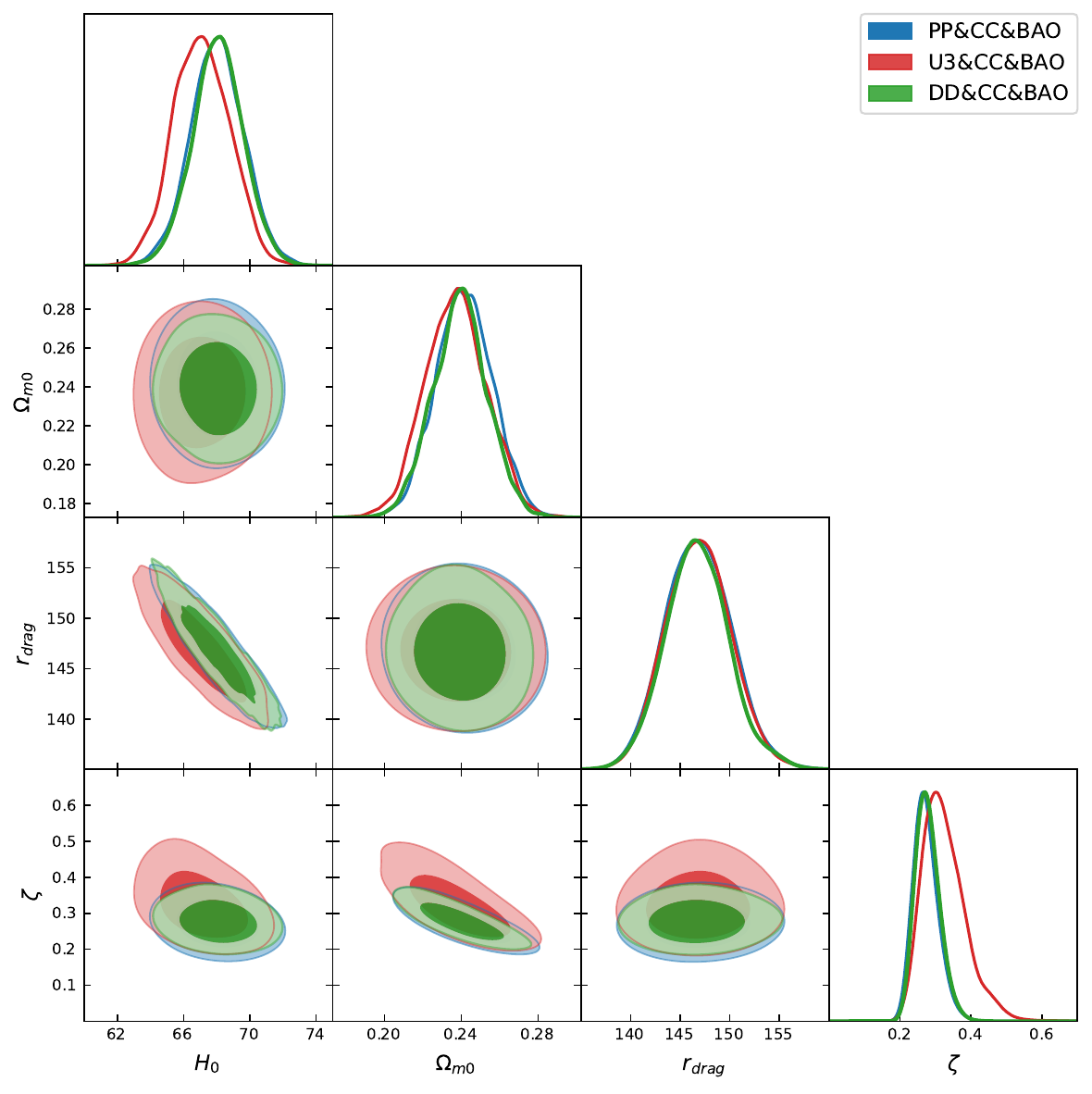}\caption{Contour
plots based on the parametric space within credible intervals for the VHDE model.}%
\label{fig1}%
\end{figure}

\begin{figure}[h]
\centering\includegraphics[width=1\textwidth]{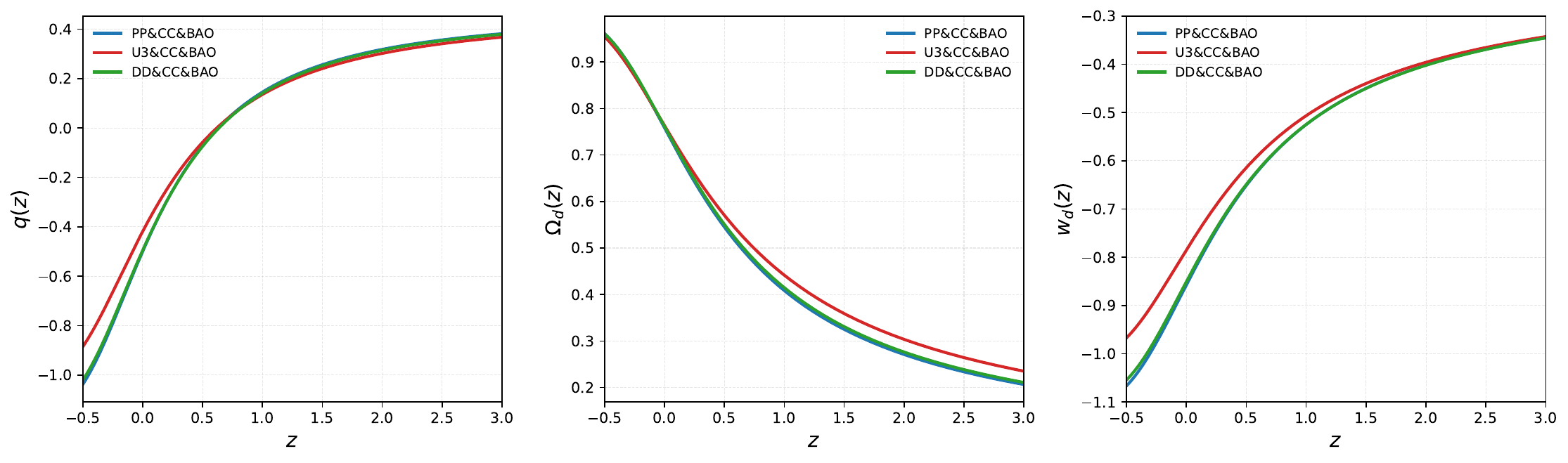}\caption{Qualitative
evolution deceleration parameter $q\left(  z\right)  $, the energy density
$\Omega_{d}\left(  z\right)  $ and the equation of state parameter
$w_{d}\left(  z\right)  $  for the posterior variables of Table \ref{tab2}.}%
\label{fig2}%
\end{figure}

Table \ref{tab2} shows the results corresponding to six distinct
configurations of the datasets described above, while in Fig. \ref{fig1} we present the confidence regions for the free parameters derived from the datasets where we combined all the different data. Furthermore, in Fig. \ref{fig2} we illustrate the qualitative evolution of the deceleration parameter $q(z)$, the $\Omega_d(z)$ and the $w_d(z)$ for the mean value of Table \ref{tab2}. The deceleration parameter $q(z)$ shows a smooth transition from the decelerated phase dominated by dust to the late-time acceleration epoch. The energy density $\Omega_d$ shows the dynamical nature of the VHDE.\\ 

The combination of the SNIa, CC, and BAO data reveals that the mean values for the Hubble parameter $H_0$ lie within the range $67.1-68.1$ $\mathrm{km/s/Mpc}$, The obtained values for the Hubble parameter are consistent with those inferred from standard late-time cosmological probes and remain close to the values typically obtained for the $\Lambda$CDM. On the other hand, the present mean value of the matter density parameter $\Omega_{m0}$ is constrained in the interval $0.236-0.242$, which are lower from that of the $\Lambda$-Cosmology. In addition, mean value for the sound horizon at the drag epoch, $r_{drag}$ is found to be in the range $143.0-146.9$, depending on the dataset combination. Finally, the mean value of parameter $\zeta=(\frac{\pi}{3} \delta^2)$ is constrained between $0.272$ and $0.327$, with the highest mean value obtained when the U3 catalogue is used. \\

Regarding the comparison of statistical indicators with respect to the $\Lambda$CDM model, we find that the VHDE model provides slightly smaller values of $\chi^2_{\min}$ for most of the dataset combinations, particularly for the PP\&BAO, U3\&BAO and DD\&BAO datasets. \\

The information criteria $\Delta\mathrm{AIC}$ remain generally positive, indicating that the VHDE model is statistically equivalent to $\Lambda$CDM. Nevertheless, from the Bayesian evidence $\Delta \ln Z$, and Jefrrey's scale, we conclude that there exist a weak evidence in support of the $\Lambda$CDM for the datasets PP\&BAO, U3\&BAO, U3\&CC\&BAO and DD\&BAO, and a strong evidence in support of the $\Lambda$CDM for the rest two datasets, that is, PP\&CC\&BAO and DD\&CC\&BAO. 

\begin{table}[tbp] \centering
\caption{Numerical outcomes inferred from the VHDE model and comparison of the statistical parameters with respect to the $\Lambda$CDM.}%
\begin{tabular}
[c]{ccccccc}\hline\hline
\textbf{Dataset} & \textbf{PP\&BAO} & \textbf{PP\&CC\&BAO} & \textbf{U3\&BAO}
& \textbf{U3\&CC\&BAO} & \textbf{DD\&BAO} & \textbf{DD\&CC\&BAO}\\\hline
$\mathbf{H}_{0}$ & $-$ & $68.0_{-1.6}^{+1.6}$ & $-$ & $67.1_{-1.7}^{+1.7}$ &
$-$ & $68.1_{-1.6}^{+1.6}$\\
$\mathbf{\Omega}_{m0}$ & $0.240_{-0.015}^{+0.015}$ & $0.242_{-0.015}^{+0.015}$
& $0.236_{-0.017}^{+0.017}$ & $0.237_{-0.016}^{+0.016}$ & $0.239_{-0.015}%
^{+0.015}$ & $0.239_{-0.014}^{+0.014}$\\
$\mathbf{r}_{drag}$ & $145.0_{-17}^{+10}$ & $146.9_{-3.4}^{+3.4}$ &
$143.0_{-17}^{+13}$ & $146.8_{-3.3}^{+3.3}$ & $145.0_{-17}^{+12}$ &
$146.7_{-3.2}^{+3.2}$\\
$\mathbf{\zeta}$ & $0.275_{-0.040}^{+0.028}$ & $0.272_{-0.040}^{+0.028}$ &
$0.327_{-0.076}^{+0.040}$ & $0.325_{-0.070}^{+0.044}$ & $0.279_{-0.040}%
^{+0.028}$ & $0.278_{-0.039}^{+0.028}$\\
$\Delta\mathbf{\chi}_{\min}^{2}$ & $-1.08$ & $+0.07$ & $-2.48$ & $-0.27$ &
$-1.03$ & $+0.17$\\
$\Delta\mathbf{AIC}$ & $+0.92$ & $+2.07$ & $-0.48$ & $+1.73$ & $+0.97$ &
$+2.17$\\
$\Delta\ln\mathbf{Z}$ & $-2.83$ & $-3.73$ & $-1.41$ & $-2.19$ & $-2.79$ &
$-3.60$\\\hline\hline
\end{tabular}
\label{tab2}
\end{table}

\section{Conclusions}
\label{sec5}
The present work dealt with the study of the VHDE model as a dark energy candidate for describing the late-time accelerated expansion of the Universe. We considered six different combinations of Type Ia supernova data, Cosmic Chronometers, and BAO measurements from the DESI DR2, in order to place constraints on the free parameters of the model, ${H_{0},\Omega_{m0},\delta}$. In addition, the parameter $r_{drag}$ was treated as a free parameter. The statistical analysis was performed with the Bayesian interface Cobaya. \\

The numerical results indicate that the mean value of the Hubble constant lies in the interval $67.1-68.1~\mathrm{km/s/Mpc}$, which is consistent with the values typically obtained for the Lambda-CDM model when using similar datasets. Furthermore, the present value of the matter density parameter is found to lie within the range $\Omega_{m0}\simeq0.236-0.242$, which is smaller than the value usually inferred in the standard $\Lambda$-cosmology. Such a lower value may have implications for the growth of cosmic structures and the $S_8$ tension. On the other hand, the VHDE parameter is constrained to $\zeta=\frac{\pi}{3}\delta^2\simeq0.272-0.327$. This interval suggests a non-negligible deviation from the standard Bekenstein-Hawking entropy, while still allowing a viable cosmological evolution. \\

Regarding the statistical comparison with the $\Lambda$CDM model, the VHDE 
model provides slightly smaller values of $\chi^2_{\min}$. However, the application of AIC suggests that these two models describe the examined datasets in a similar way, mainly because the VHDE scenario introduces one additional free parameter. However, the application of Jeffrey's scale for Bayesian evidence reveals that the data have a moderate preference in favor of the $\Lambda$CDM model. \\

We continue our discussion with the comparison of our results with those of previous entropic modified cosmological models, where the same datasets were applied. In Kaniadakis HDE \cite{Luciano1}, the median value for the matter density was found to be $\Omega_{m0}\simeq0.260-0.273$, and for the Barrow–Tsallis entropic model \cite{tsallis01} within the range $\Omega_{m0}\simeq0.280-0.298$. Furthermore, for the recent Luciano–Saridakis entropic theory \cite{Leizerovich:2026pfy}, the energy density of matter was found to be $\Omega_{m0}\simeq0.290$. These values are larger than those obtained from the VHDE model. This difference can be seen as a result of the use of the different definitions of the VHDE model as discussed above. Moreover, the smaller value obtained for $\Omega_{m0}$ in the VHDE may affect the $S_8$ value of the growth of matter perturbations, which means that such a study should be explored in the future. Finally, regarding the comparison of the statistical parameters, all models fit the data in a similar way and are statistically indistinguishable. \\

We find that the VHDE model is consistent with current late-time cosmological observations and provides a viable framework for describing the late-time acceleration of the universe. In a future work, we shall focus on extending the present analysis to the study of cosmological perturbations.

\begin{acknowledgments}
AP thanks the support of VRIDT through Resoluci\'{o}n VRIDT No. 096/2022 and
Resoluci\'{o}n VRIDT No. 021/2026. Part of this study was supported by
FONDECYT 1240514.
\end{acknowledgments}


\end{document}